\documentclass[11pt]{article}
\usepackage[margin=1in]{geometry}
\usepackage{array,booktabs,amsmath}
\usepackage{graphicx}
\usepackage[table]{xcolor}
\usepackage{tikz}
\definecolor{tablehead}{HTML}{E8F1F5}
\definecolor{tablerule}{HTML}{8FA7B5}
\definecolor{tablerow}{HTML}{F5F8FA}
\definecolor{scoreteal}{HTML}{087F83}
\usetikzlibrary{arrows.meta,positioning}
\definecolor{spotblue}{HTML}{2563A6}
\definecolor{spotteal}{HTML}{087F80}
\definecolor{spotamber}{HTML}{A66A08}
\definecolor{spotink}{HTML}{243449}

\usepackage[hidelinks]{hyperref}
\usepackage[expansion=false]{microtype}
\title{Spotlights: Discovering Improvement Opportunities\\in Software Repositories}\author{\normalsize Udi Barzelay, Ophir Azulai, Idan Friedman, Inbar Shapira\\
\normalsize Foad Abo Dahood, Yevgeny Burshtein, Orit Prince, Michael Soloveitchik\\
\normalsize Oshri Naparstek, Roi Pony, Tal Drory, Michael Factor}
\date{}
\begin{document}
\maketitle

\begin{abstract}
Coding agents and evolutionary code-search systems can improve implementations
once a target and evaluation criterion have been specified. Applying these
methods to an existing software repository raises an earlier question: which
implementation choices are worth investigating for a high-level engineering
objective? We introduce \emph{optimization-opportunity discovery}, the
repository-level task of identifying candidate source regions, explaining
how they relate to the objective, and proposing possible changes. The task
takes as input a repository, an engineering objective, and optional runtime
evidence such as offline telemetry observations or profiles. It does not
require the user to specify a defect, bottleneck, or code location.
We present \emph{Spotlights}, a system that performs this task through logical
repository mapping, successive agent reviews, and optional research linking
candidates to relevant techniques. We evaluate Spotlights across model serving,
document retrieval, blockchain ordering, and document processing. Across three
cases, it recovers seven of nine expert-selected targets. In the reliability
study, 70\% of the top ten candidates meet the stated correctness and severity
thresholds. Across five repeated retrieval runs, 73.6\% of candidate occurrences
have a matching source region in all five runs. Spotlights also rediscovers
the target of a withheld retrieval optimization and connects it to a relevant
tiling technique. In an implementation study, a discovered change reduces
end-to-end page-processing runtime by 10.6\% while preserving measured output
quality. These results establish optimization-opportunity discovery as a
distinct and empirically evaluable step between a broad engineering objective
and subsequent implementation and validation.
\end{abstract}

\section{Introduction}

Automated search is becoming a productive way to improve algorithms and
software implementations. FunSearch combines language-model generation
with evaluation to discover programs for a supplied problem
specification~\cite{funsearch}. AlphaEvolve extends evolutionary code search
to scientific and engineering tasks~\cite{alphaevolve}, and OpenEvolve
provides an open implementation of a similar search workflow~\cite{openevolve}.
These searches evaluate program variants against a task and an evaluation
criterion supplied in advance. Applying them to an existing repository
raises an earlier question: which implementation choices offer worthwhile
opportunities for improvement?

That task begins with a different kind of input. An engineer may have a
repository and a goal concerning efficiency, resource use, or reliability,
but no list of changes to attempt. The repository may contain many plausible
starting points: an algorithm, a representation, a scheduling policy, a
validation rule, or a configuration decision. A useful discovery connects
one of these choices to the objective and explains an alternative worth
examining. Producing such a connection requires understanding both the code's
role and the circumstances under which a change could matter.

We call this task \emph{optimization-opportunity discovery}. Its output is
a set of candidate source regions, each accompanied by an explanation of the
opportunity and proposed changes. Making these candidates explicit
provides a concrete object of study: we can assess whether the reasoning is
sound, whether known opportunities are recovered, and whether independent
runs repeatedly identify the same regions. Candidates also give developers
and automated tools specific starting points for subsequent investigation.

We present \textbf{Spotlights}, a system that connects repository analysis
with an engineering objective to identify and explain opportunities for
improvement. Its agents use a logical module map to guide source inspection,
review candidate opportunities, and develop proposed changes.
Optional technical research connects these proposed changes to relevant
algorithms and implementation techniques.

We applied Spotlights to objectives including reducing serving latency
and improving the reliability of a blockchain ordering service. In each
application, the objective guides the search for relevant decisions in the
code; the candidate locations and proposed changes emerge from that
search.

Across these applications, Spotlights identifies decisions governing cache
reuse, request routing, and transaction validation. The evaluation recovers
expert-selected locations and finds recurring source regions across repeated
runs. In document retrieval, the system identifies the computation targeted
by a known optimization and links it to a relevant memory-saving technique
while the solution is withheld from its inputs. Finally, implementing one
proposal in a document-processing pipeline reduces runtime while preserving
measured output quality. Together, these results show how discovery turns
a broad objective into specific, technically motivated starting points
for engineering work.

The paper makes three contributions:
\begin{itemize}
\item A formulation of repository-level opportunity discovery with candidate
      source regions, rationales, and proposed changes as its output.
\item The Spotlights system for discovering and developing actionable
      improvement opportunities from repository context.
\item An empirical evaluation of candidate quality, recovery of known
      opportunities, stability, and cost, together with an implementation
      case demonstrating practical utility.
\end{itemize}

\section{Related Work}

\paragraph{Program search and automated improvement.}
FunSearch~\cite{funsearch} and AlphaEvolve~\cite{alphaevolve} demonstrate
algorithmic discovery through repeated program generation and evaluation.
OpenEvolve~\cite{openevolve} exposes evolutionary search through an initial
program and evaluator, while CORAL~\cite{coral} studies multi-agent
evolution for discovery. These systems motivate the value of identifying
problems worth investigating in an existing repository. Spotlights makes
candidate discovery explicit and evaluates the locations and opportunities
it produces. The IOCR experiment illustrates one use of a candidate through
CORAL; the discovery output can also support direct investigation by a developer.

\paragraph{Repository analysis.}
Agentless~\cite{agentless} localizes and repairs code using an issue
description, and RepoAudit~\cite{repoaudit} investigates repository-level
code auditing. Such work demonstrates the use of language-model agents
to connect source analysis with a specified task. Spotlights conditions
the search on an engineering objective and produces opportunities expressed
as source locations, rationales, and proposed changes.

\paragraph{Optimization-target selection.}
POLO~\cite{polo}, PerfAgent~\cite{perfagent}, and agent-agnostic application
optimization~\cite{agentagnosticperf} use runtime evidence in their
optimization workflows. EffiHolmes~\cite{effiholmes} uses differential
profiles to localize repair sites for reproducible inefficiencies.
Beyond Local Code Optimization~\cite{beyondlocal} uses structural and
architectural evidence to identify and prioritize cross-component
opportunities. These approaches are relevant comparisons for discovery:
they use different inputs and analysis scopes to identify where to act.
Spotlights studies discovery from a high-level engineering objective, with
workload or runtime context supplied when available. Its evaluation focuses
on the quality and recurrence of the nominated opportunities.

\section{The Opportunity-Discovery Problem}
\label{sec:task}

Let $R$ be a repository snapshot, $o$ an engineering objective, $h$ optional
context, and $S$ an analysis scope. Discovery returns a collection of
candidates,
\[
 D(R,o,h,S) \longrightarrow C.
\]
The user supplies the repository and an objective stating a desired
improvement at the system level. Optional context describes conditions
under which that improvement matters, such as a workload description or
offline telemetry: measurements of system behavior supplied as context. The scope can cover the repository or a selection of modules
from the map constructed by Spotlights. Thus, the user specifies the
engineering question and search scope without enumerating the candidate
locations.

A \emph{candidate} is an opportunity for improvement, described by its source
locations, current implementation, rationale, and estimated impact. A
\emph{proposal} describes a possible change addressing that opportunity;
a candidate can have several proposals. A location identifies a file and line interval; the general
schema supports multiple locations. The rationale explains why that region
is relevant to the objective. A proposal may draw on a technique from a
paper, documentation, or another implementation. The impact label is a model-assigned rating of expected
importance: high, medium, or low. An accompanying explanation identifies
the workload-level measure that could improve and why the region offers
room for improvement.

Section~\ref{sec:example} illustrates this output with a candidate from one
of the experimental runs.

\section{Spotlights}
\label{sec:system}

\begin{figure}[htbp]
\centering
\begin{tikzpicture}
\node[anchor=south west,inner sep=0] (overview) at (0,0)
  {\includegraphics[width=\linewidth]{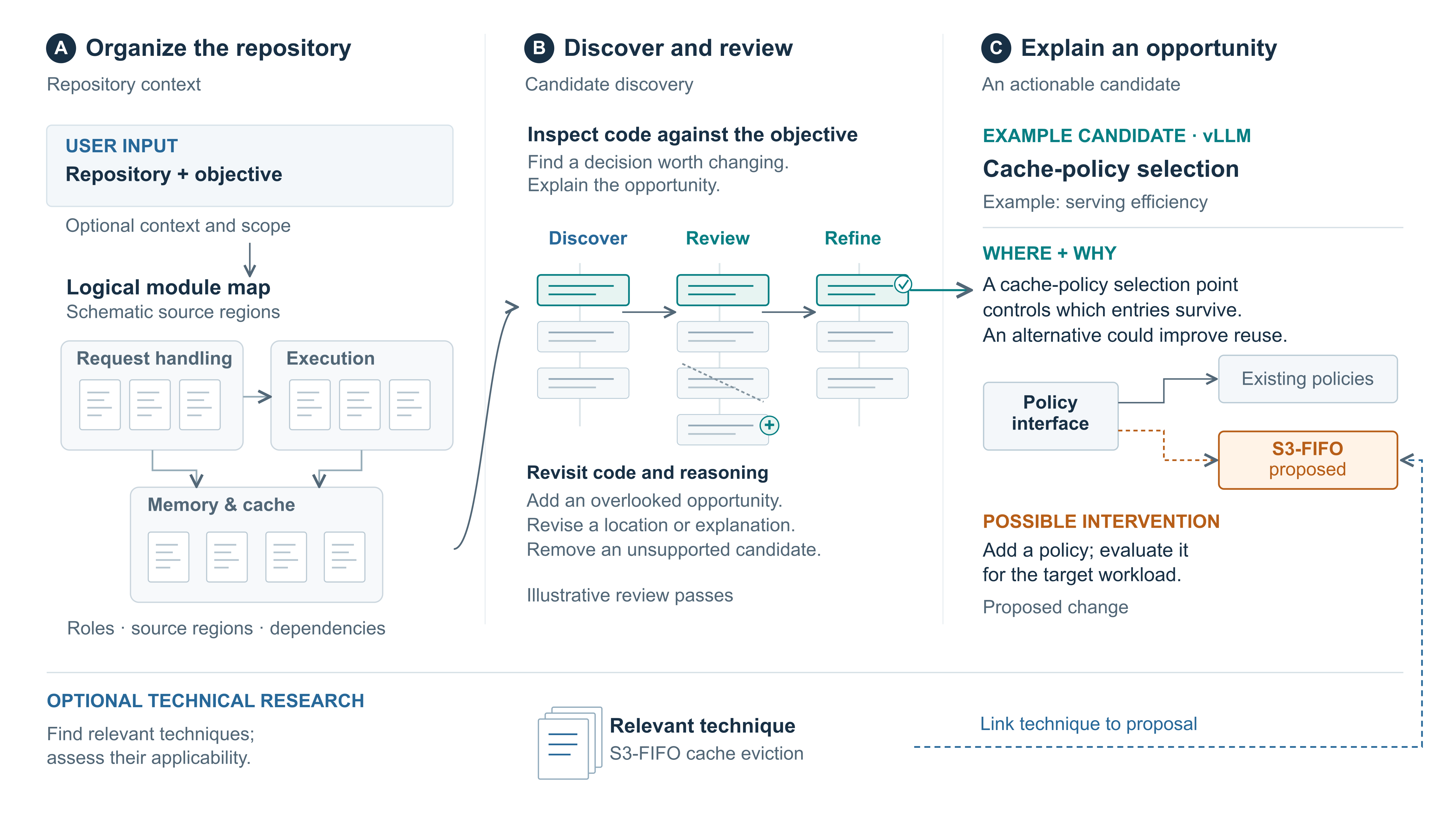}};
\begin{scope}[x={(overview.south east)},y={(overview.north west)}]
\fill[white] (0.040,0.708) rectangle (0.315,0.740);
\node[anchor=west,inner sep=0,font=\sffamily\fontsize{6}{7}\selectfont,
  text=spotink] at (0.044,0.726) {Optional: workload + offline telemetry};
\fill[white] (0.672,0.349) rectangle (0.960,0.391);
\node[anchor=west,inner sep=0,font=\sffamily\bfseries\fontsize{7}{8}\selectfont,
  text=spotamber] at (0.675,0.371) {PROPOSAL};
\end{scope}
\end{tikzpicture}
\caption{Spotlights connects a repository and an engineering objective to
candidate improvement opportunities. Optional workload descriptions and
offline telemetry observations provide operating context for discovery.
Panel A organizes the code into logical modules; panel B shows candidates
being added, revised, or removed during review; panel C explains one
candidate and its proposal. The module map and review passes are schematic. The example concerns vLLM, a language-model serving system~\cite{vllm_background},
and S3-FIFO, an alternative policy for deciding which cache entries to
retain~\cite{s3fifo}. Section~\ref{sec:example} develops this example.
Optional technical research connects techniques to proposed changes.}
\label{fig:system}
\end{figure}
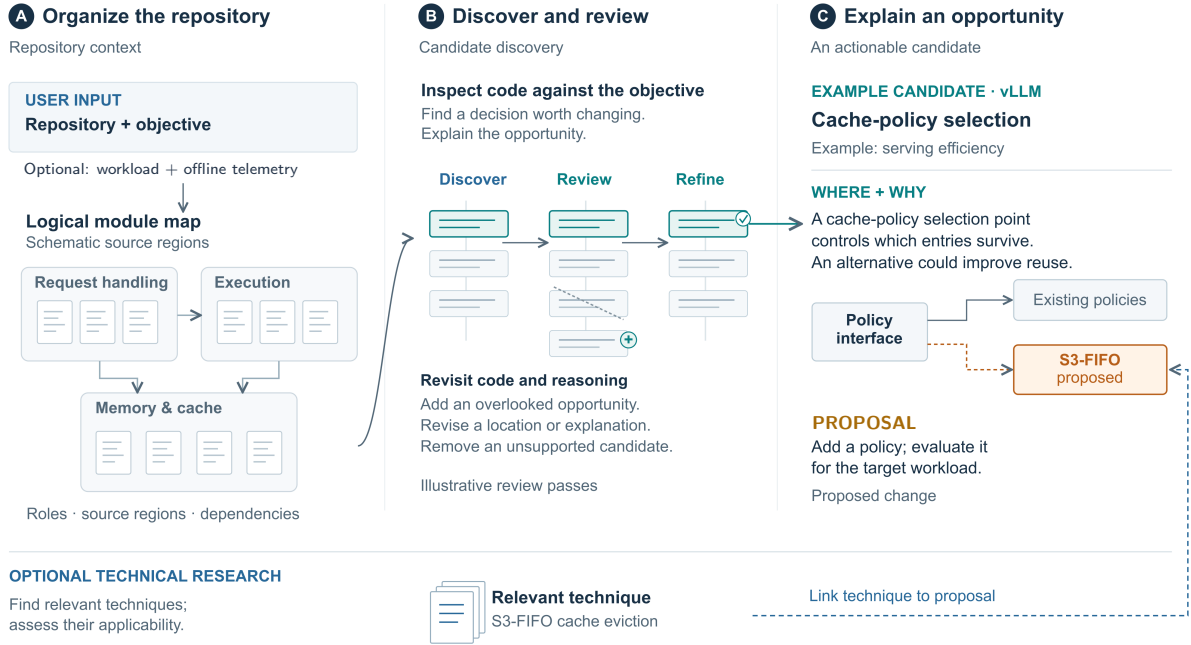

Figure~\ref{fig:system} outlines the discovery process; the following
subsections describe how each operation produces and refines its output.

\subsection{Logical Repository Mapping}

A repository's directory structure provides an initial organization, but
analysis also requires understanding what its components do. A mapping agent
examines directories, imports, manifests, documentation, and representative
source files to construct logical modules: groups of code with related
responsibilities that serve as the units of analysis. The map records each
module's role, path, representative files, and dependencies. These fields
give the discovery agent starting points for reading the code and context
for interpreting its behavior. Modules may contain submodules; analysis is performed on the selected
modules at the finest level of the map.

The map is independent of the engineering objective, allowing reuse across
goals. A caller can select modules from the completed map or analyze the
repository without a module filter. Candidate generation operates within the selected modules.

\subsection{Candidate Discovery and Iterative Review}

For each selected module, a coding agent receives its map entry, repository
context, the engineering objective, and any optional context. It starts from
the representative files and inspects relevant source code to identify
choices that could be changed. The discovery instructions ask for a
plausible improvement mechanism, a specific source location, and a way to
check the behavior of a changed implementation, such as existing tests or
expected outputs. The agent records each opportunity using the candidate schema in
Section~\ref{sec:task}.

Spotlights alternates coding agents for a configured number of review
passes. Each agent examines the source and the preceding agent's candidate
list, and can add an overlooked opportunity, revise a location or rationale,
or remove a candidate.

The orchestration layer checks record fields, file membership, and line
ranges, and preserves identifiers across revisions. These checks keep the
candidate descriptions linked to valid source regions. Models, reasoning
settings, budgets, and review count are configurable.

\subsection{Proposals and Technical Research}

Proposal agents use model knowledge to suggest changes for each candidate. An optional
research stage also searches papers, documentation, and related
implementations for techniques relevant to the module and objective.

The research stage collects source-backed descriptions of relevant algorithms
and implementation techniques. For each candidate and retrieved technique,
a proposal agent examines whether the technique applies at the candidate
location and describes the change needed to use it there. The resulting
source-attributed proposal links the proposed change to its supporting source. 

Candidates and proposals are exposed as structured JSON and a readable
report. Within each module, the report generator sorts candidates by
descending number of source-attributed proposals and then by identifier.
This rule determines presentation within a module. The experimental
rankings used to measure recovery and quality are described separately in
Section~\ref{sec:protocol}.

\section{Experimental Setup}
\label{sec:studies}
\label{sec:cases}

We first describe the applications and run inputs, then the evaluation
protocols. Appendices~\ref{app:quality-data}--\ref{app:iocr-details}
provide candidate-level scores, reference matches, repeated-run measurements
and implementation details.

\subsection{Applications and Input Conditions}

Each application supplies a repository, an engineering objective, and an
analysis scope, with additional context where available.
The applications cover language-model serving (vLLM and llm-d), document
retrieval (ColPali), blockchain ordering (Fabric-X), and document processing
(IOCR). Table~\ref{tab:cases} summarizes the principal runs. The recall evaluation
also includes a separate vLLM run and a Workload Variant Autoscaler run,
described below.

\begin{table}[htbp]
  \centering
  \small
  \renewcommand{\arraystretch}{1.18}
\arrayrulecolor{tablerule}
\rowcolors{2}{tablerow}{white}
\begin{tabular}{@{}>{\raggedright\arraybackslash}p{0.20\linewidth}
                  >{\raggedright\arraybackslash}p{0.23\linewidth}
                  >{\raggedright\arraybackslash}p{0.24\linewidth}
                  >{\raggedright\arraybackslash}p{0.23\linewidth}@{}}
    \toprule
\rowcolor{tablehead}
    Subject & Objective & Analysis scope & Additional context \\
    \midrule
    vLLM subset & Serving latency & 8 of 59 modules & Repeated interactions with an agent \\
    \addlinespace
    vLLM cache offload & Serving latency & Cache-offload subsystem & Offline hit-rate anomaly \\
    \addlinespace
    ColPali & Retrieval latency and throughput & Retrieval-engine package & None \\
    \addlinespace
    llm-d router & Serving latency & Repository run & None \\
    \addlinespace
    Fabric-X orderer & Reliability and fault tolerance & 30 modules & None \\
    \addlinespace
    IOCR & End-to-end page latency & No module filter & None \\
    \bottomrule
  \end{tabular}
  \caption{Objectives and input conditions of the principal case studies.
  The ColPali runs used a user-selected retrieval-engine package; IOCR used repository-wide
  analysis, with completed candidate output from 17 modules.}
  \label{tab:cases}
\end{table}
\paragraph{Serving and resource management.}
vLLM is a system for serving language models~\cite{vllm_background}.
It caches attention keys and values computed for previous tokens so they
can be reused during generation. CPU offloading stores part of this cache
in host memory; an eviction policy determines which entries to retain.
The vLLM subset run analyzes eight of the 59 finest-grained modules in its logical
map, with an objective of reducing serving latency and a workload hint
describing multi-turn interactions with an agent. We use this run to
illustrate candidate output and study how successive reviews change it.
A second run, the vLLM cache-offload case, supplies an offline observation
about cache hit rate---how often requested data is found in the cache---as
additional context for analysis of that subsystem. A third, the vLLM recall
run, tests recovery of an expert-selected cache-policy dispatch location.
Appendix~\ref{app:execution} gives the subset run's module-selection and
agent configurations.

\paragraph{Request routing.}
The llm-d router selects servers to process language-model
requests~\cite{llmd_router}. Request processing includes prefill, which
processes the input tokens, and decode, which generates output tokens.
The router can direct these phases to separate serving resources and
choose among servers using scores derived from their state. We give
Spotlights a serving-latency objective on the router repository and compare
its discoveries with seven expert-selected locations. This case tests
whether discovery reaches decisions governing how requests are distributed.

\paragraph{Resource scaling.}
The Workload Variant Autoscaler (WVA) adjusts the number of model-server
replicas as demand changes~\cite{wva_project}. Its objective is to reduce GPU cost while
meeting latency and throughput requirements. The reference target is the
saturation analyzer, which uses cache utilization and queue length to
estimate available capacity and decide whether to add or remove replicas.
The WVA run tests whether Spotlights identifies this scaling logic as an
opportunity for improvement.

\paragraph{Document retrieval.}
ColPali retrieves document pages relevant to a text query by comparing
vector representations of the query and page
images~\cite{colpali_background}. Its multi-vector representation supports
fine-grained matching between query tokens and page content. We apply
Spotlights to the user-selected retrieval-engine package, asking it to
reduce query-processing time and increase throughput.

This application provides a known opportunity against which to test
discovery. Our FlashMaxSim work optimizes the MaxSim computation used to
score query--document similarity~\cite{flashmaxsim}. We withhold that
solution from Spotlights and test whether it identifies the relevant
computation and a useful change. Five runs under the same inputs
also let us examine whether the system repeatedly discovers the same
source regions.

\paragraph{Reliability.}
The Hyperledger Fabric-X ordering service establishes a common order of
transactions and groups them into blocks~\cite{fabricx_orderer}. The study
seeks reliability and fault-tolerance improvements under internal failures
and malicious clients. Its objective includes crash faults, where a party
stops operating, and Byzantine faults, where a party may deviate
arbitrarily from the protocol. The requested fault model allows up to $F$
faulty parties among $N=3F+1$ participants. We analyze 30 modules and assess
whether the discovered opportunities address the correctness and
robustness of this ordering process.

\paragraph{Document processing.}
IOCR is an industrial optical character recognition (OCR) pipeline that
converts page images into structured text. Its processing includes a
word-detection model that locates text regions in a page. We run Spotlights
with a CPU-only, end-to-end page-processing latency objective and no module
filter. We then follow one discovered opportunity through implementation
and measurement to examine how its location and proposal guide a concrete
change to the pipeline.

\subsection{Evaluation Protocol}
\label{sec:protocol}

\paragraph{Candidate quality.}
The Fabric-X assessment considers whether an opportunity is technically
correct and how consequential the identified issue is. A listwise
judge evaluates the candidate list jointly to produce an order; correctness and severity
scores on a 1--10 scale are assigned to its top 20 candidates.
We count a candidate as a hit when correctness is at least 7 and severity
is at least 5. Precision@k is the fraction of the first $k$ candidates
meeting both thresholds; we report it for $k=10$ and $k=20$.

\paragraph{Recovery of expert-selected targets.}
For the vLLM, llm-d router, and WVA recall runs, a candidate matches an
expert reference when its line span overlaps the reference in the same
file, with no additional tolerance. Overall recall counts matches anywhere
in the output. Recall under a review budget counts only matches among the
first $k$ candidates in that run's evaluation order. A separate listwise
judge ranks the llm-d router candidates jointly, evaluating the candidate
list as a whole to produce a single order.

\paragraph{Rediscovery and repeatability.}
For the ColPali rediscovery test, the experiment specifies a retrieval
cutoff of January 1, 2026, preceding the FlashMaxSim publication.
The runs use the same repository commit, Spotlights revision, objective,
module filter, and prompt set. Each run generates its own module map and
candidates. We measure location recurrence using overlapping spans in
the same file, both over all candidate occurrences and over source coverage
analyzed in every run. An occurrence is one candidate in one run. For
impact-label stability, we connect candidates from different runs when
their source spans overlap and form groups from candidates linked directly
or through other candidates. We compare labels in groups represented in
all five runs.

\paragraph{Review activity and cost.}
In the vLLM subset, we count candidates added or revised during review and
candidates with source-attributed proposals. For ColPali and IOCR, usage and
elapsed time quantify the cost of discovery. Contracted rates are the
model-service prices used for these runs; public list rates refer to the
providers' published prices. These figures cover Spotlights
runs and exclude subsequent optimization and benchmarking.
Appendix~\ref{app:execution} gives accounting details.

\paragraph{Implementation study.}
We select one IOCR candidate and use its location and proposal to define
an implementation task for CORAL, a framework for program
optimization~\cite{coral}. The resulting modified program is compared with
the baseline on the same set of page images. Each timed repetition covers
a complete end-to-end pass over that set. Baseline and modified versions
run back-to-back on the same node, with four allocated CPU cores, one
warm-up, and five timed repetitions per
version. Appendix~\ref{app:iocr-details} specifies the implementation
and measurement details. Output quality is measured by
character accuracy against reference text, averaged over the images.

\section{Results}

\subsection{An Illustrative Discovery}
\label{sec:example}

\begin{figure}[htbp]
\centering
\includegraphics[width=\linewidth]{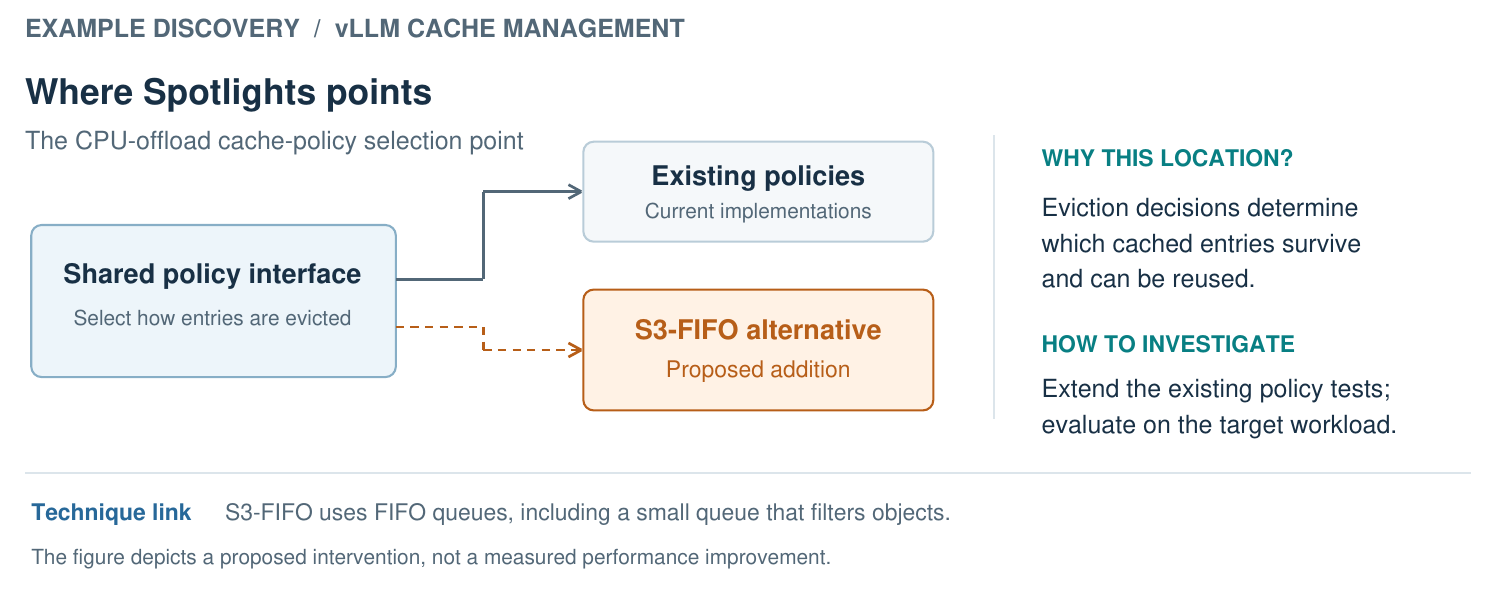}
\caption{A discovered opportunity in vLLM: the shared cache-policy interface
provides a place to add and evaluate an alternative eviction strategy.
Spotlights links the proposed S3-FIFO policy to a published technique~\cite{s3fifo}.
}
\label{fig:record}
\end{figure}

To illustrate the candidates produced by Spotlights, we show one example
from the vLLM subset run described in Section~\ref{sec:studies}.
Spotlights identifies the registration table that selects the CPU cache
eviction policy and proposes an additional policy, S3-FIFO. This published
cache-eviction algorithm uses first-in-first-out queues, including a small
queue that filters objects before entry into the main cache~\cite{s3fifo}.
Figure~\ref{fig:record} shows how this candidate connects the serving
objective to an extension point in the implementation. The separate
vLLM cache-offload run also identifies an eviction-policy change,
using an offline cache-hit-rate anomaly as additional context.

\subsection{Are the Discovered Candidates Technically Meaningful?}
\label{sec:precision}
The Fabric-X run returns 119 candidates. Under the quality criterion in
Section~\ref{sec:protocol}, precision@10 is $7/10=0.70$ and precision@20 is
$9/20=0.45$. The first-ranked candidate concerns
detection of a participant signing conflicting versions of a batch. Its
proposal calls for counting each signer once and requiring signed evidence
from the participant accused of issuing conflicting batches. It scores 10 on
both dimensions.

The fifth-ranked candidate concerns the rules specifying
whose signatures authorize a block. Its proposal checks that the signature
sub-rules reference every distinct authorized participant, strengthening
the check on the identity list. It scores 10 on correctness and 7 on
severity. These candidates connect the reliability objective to concrete
decisions about detecting inconsistent behavior and accepting block policies.

Appendix~\ref{app:quality-data} lists all 20 candidate scores.
The threshold matters: three of the nine hits have severity exactly 5,
while two excluded candidates have correctness 10 and severity 2. The distinction between correctness and severity therefore affects which
opportunities merit attention under this evaluation criterion.

\subsection{Does Discovery Recover Expert-Selected Targets?}
\label{sec:recall}
\begin{figure}[htbp]
\centering
\includegraphics[width=\linewidth]{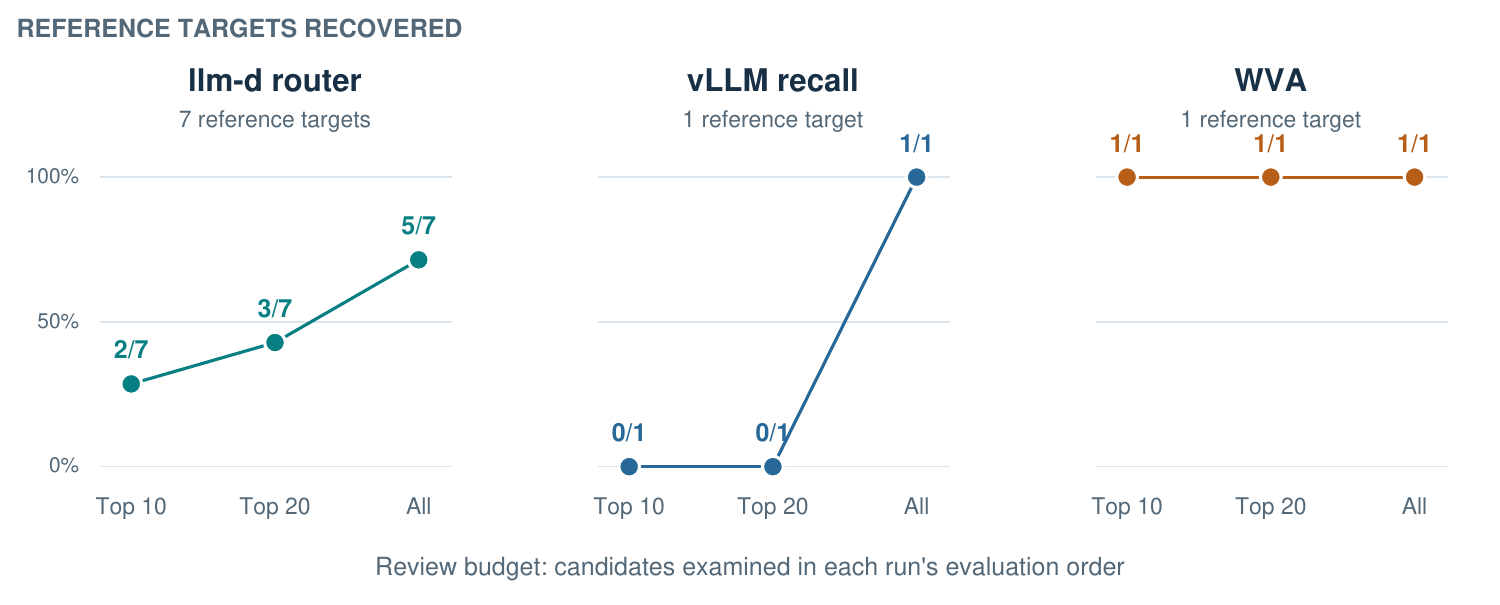}
\caption{Recovery under increasing review budgets. Labels give recovered
targets divided by the size of the reference set. Each run uses its own
evaluation order; ``All'' includes its full output. The vLLM recall run is
distinct from the 105-candidate subset example.}
\label{fig:recall-budget}
\end{figure}

Figure~\ref{fig:recall-budget} reports recovery of expert-selected targets using
the matching rule and review budgets defined in Section~\ref{sec:protocol}.
Appendix~\ref{app:quality-data} reports the result for each reference target.

The llm-d router run recovers five of seven references, including two in the
top ten and three in the top twenty, from an output of 100 candidates.
The rank-four match concerns the threshold deciding whether to separate
prefill and decode across serving resources. This decision trades the
benefit of separating the work against the cost of transferring cached
state. The rank-nine match concerns the weights combining server scores
for request routing: changing them alters how cache availability and load
influence the selected destination. Both matches identify decisions that
control the distribution of serving work.

The two smaller reference sets each contain one target. In the separate
261-candidate vLLM recall run,
the cache-manager selection logic is recovered at rank 57. This location
selects a CPU cache manager according to the configured eviction policy,
providing a point at which that choice can be changed. In WVA, the
saturation analyzer appears at rank four, locating the logic that converts
observed cache utilization and queue capacity into scaling decisions. These cases
show why target recovery and review order should be measured separately:
a relevant location can be found but still require substantial review effort.

\subsection{Rediscovering a Known Retrieval Optimization}
\label{sec:rediscovery}

The ColPali runs nominate MaxSim, the computation that scores how well a
document matches a query. For each query vector, MaxSim finds the most
similar document vector, then sums these similarities. A direct
implementation forms the pairwise similarity array before reducing it to
a score. This is the computation optimized by
FlashMaxSim~\cite{flashmaxsim}.

Spotlights connects a tiled implementation to
FlashAttention, an attention algorithm that reduces memory transfers by
computing in small blocks, or tiles~\cite{flashattention}. The proposal
computes similarities block by block and updates the running maxima,
avoiding storage of the full intermediate array. It proposes checking
numerical agreement with the original implementation. The outputs do not name
FlashMaxSim; they identify its target computation and connect it to a
relevant technique.

\subsection{Stability Across Repeated Runs}
\begin{figure}[htbp]
\centering
\includegraphics[width=\linewidth]{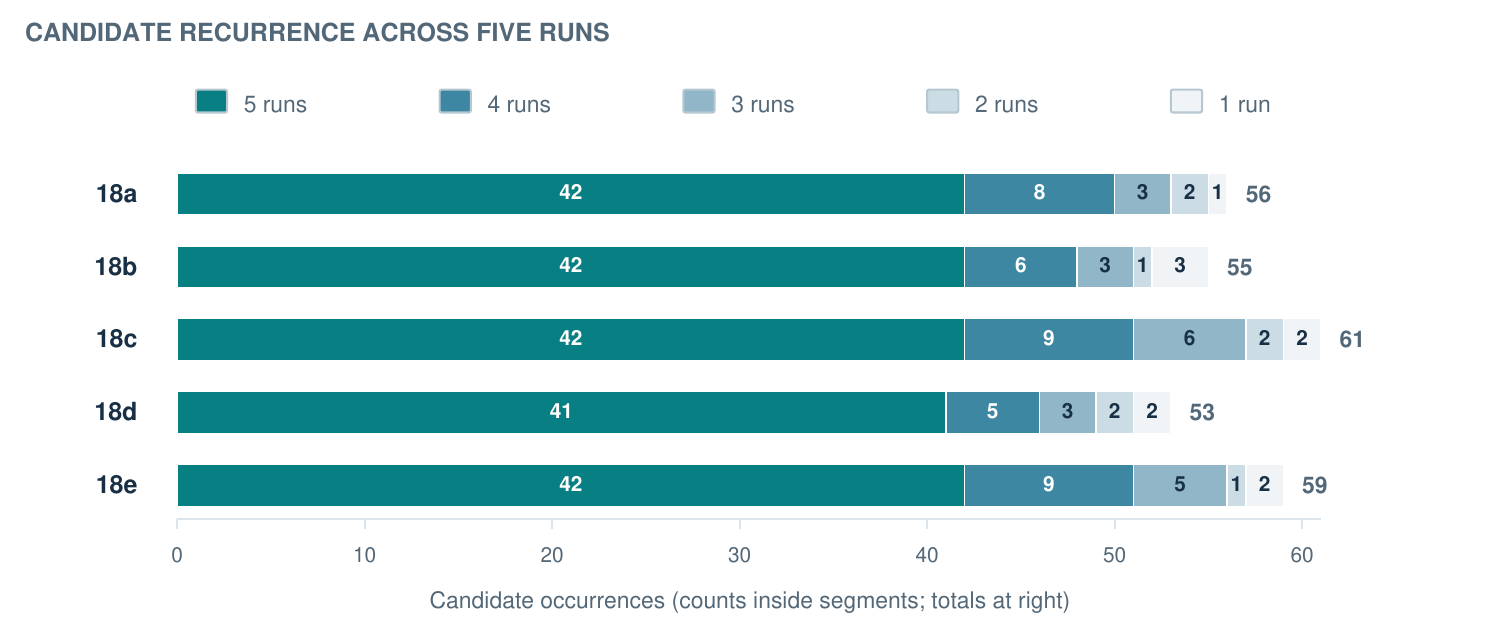}
\caption{Candidate recurrence in five ColPali runs with identical inputs,
labeled 18a--18e. Each segment counts candidates whose source regions directly overlap a
candidate in the indicated number of runs, including their own run.
Numbers inside segments are counts; numbers at right are per-run totals.}
\label{fig:stability}
\end{figure}

Across five ColPali runs, 209 of 284 candidate occurrences (73.6\%) overlap
a candidate in every other run under the rule in Section~\ref{sec:protocol}.
Figure~\ref{fig:stability} shows the per-run distribution, indicating a
substantial recurring set of source regions despite variation in the
remaining candidates.

Module maps vary in granularity and coverage. Restricting the calculation
to source coverage analyzed in all five runs leaves 263 candidate
occurrences and yields $209/263=79.5\%$. This restriction excludes 21
occurrences outside the shared coverage; all 209 occurrences matched
across every run remain in the calculation.
The difference between these estimates shows that variation in module
coverage contributes to variation in the candidate set.

Of the 41 groups of overlapping candidates present in all five runs, 15 (36.6\%) have unanimous impact labels, and 34 (82.9\%) have a label
shared by at least four runs. Thus, recurring source locations can receive different impact ratings
across runs.

\subsection{How Do Review and Research Shape the Output?}
\label{sec:activity}
The vLLM subset produces 105 candidates across 58 files. Review passes
add 32 of the final candidates and revise 78. Source-attributed proposals cover 46 candidates (43.8\%). Thus, the final candidate set develops substantially through review, and
external-source proposals cover a subset of the discovered locations. The
remaining candidates can still carry proposals generated from model knowledge.

\subsection{What Does Discovery Cost?}
The five ColPali runs cost \$131.28--\$164.66 at contracted rates, consume
approximately 84--100 million tokens, and take 4.95--5.30 hours. The IOCR
discovery run costs \$123.60 at contracted rates (\$177.66 at public list
rates), consumes 94.7 million tokens, and takes approximately 4.3 hours.

Across the ColPali runs, candidate discovery accounts for 3.9\% of stage
cost, approximately \$5.75 per run. Source-attributed proposal generation
accounts for 69.3\%, proposals from model knowledge for 16.2\%, and
external-source retrieval for 10.7\%. This breakdown excludes module-map construction;
percentages exceed 100\% slightly because of rounding. Most expenditure supports the development of proposed changes and their
connections to sources. This distribution identifies a concrete design
tradeoff: how much research to perform for each candidate, and which
proposals justify that investment.

\subsection{From a Proposal to a Measured Improvement}
\label{sec:iocr}
The IOCR run returns 126 candidates over 17 modules. The selected,
fourth-ranked candidate identifies the model-loading step, where the
detector configures ONNX Runtime, the software that executes
the word-detection model~\cite{onnxruntime_threading}. Its session holds the loaded model and execution
settings. The proposal enables a CPU memory arena, which pools memory for
reuse, and revises thread settings to reduce allocation and scheduling
overhead.

The implementation applies these settings and updates the code that
configures the session. We evaluate the modified pipeline under the
protocol in Section~\ref{sec:protocol}.

Median runtime falls from 150.970 to 135.026 seconds, a 10.6\% reduction.
Mean character accuracy is 95.53\% for both versions. The runtime ratio is
0.8944, with a two-standard-error band of [0.8818, 0.9070]. The discovery
thus supplied a specific target and change that led to an implemented
and measured improvement in the pipeline.

\section{Discussion}

\paragraph{Discovery as an object of evaluation.}
Candidate quality, visibility in the review order, and repeatability are
distinct properties. A technically relevant candidate may appear late in
the list, while a recurring location may receive different impact ratings.
Assessing these properties before implementation helps developers decide
which opportunities warrant further investigation.

\paragraph{Finding and prioritizing opportunities.}
Review order determines how much effort an engineer spends finding a useful
candidate. Prioritization should therefore consider both the expected value
of a change and the work needed to investigate it. Candidate
significance, source support, and implementation effort provide different
signals for this decision; comparing them with expert judgments and
implementation outcomes is a concrete direction for improving discovery.

\section{Future Work}

Future work will extend the use of telemetry from supplied offline
observations to automated collection and interpretation of runtime signals,
so discovery can respond to changes in workloads and operating conditions.
Impact assessment will compare predicted importance with expert judgments
and measured outcomes, accounting for the effort and extent of code changes
needed to investigate each opportunity. Cost reduction will focus on
selective technical research and adaptive review budgets, evaluating how
these choices affect the quality of the resulting candidates.

\section{Conclusion}

Spotlights turns a broad engineering objective into concrete opportunities
for improving an existing software repository. Logical mapping gives agents
a structured basis for source inspection; iterative review refines candidate
locations and reasoning; and optional technical research connects proposed
changes to supporting sources. Each candidate explains where to investigate
and why, giving developers a concrete starting point for further work.

Across serving, retrieval, blockchain ordering, and document processing,
the evaluation identifies technically meaningful opportunities, recovers
expert-selected targets, and finds recurring source regions across runs.
The retrieval study connects a known optimization target to a relevant
published technique, while the document-processing case follows a proposal
through implementation to a measured improvement. These results demonstrate
the value of making opportunity discovery an explicit, inspectable step in
software improvement.

\bibliographystyle{plain}
\small
\bibliography{references}
\normalsize
\appendix
\section{Additional Experimental Details}
\label{app:execution}

\paragraph{Module selection.}
In the vLLM subset configuration, selecting a parent module includes its
finest-grained descendant modules in the analysis.

\paragraph{Review sequence and proposals.}
The vLLM subset uses a Claude initialization pass followed by Codex, Claude,
and Codex reviews. Initialization produces 76 candidates, three of which
are removed. The passes make 149 modifications in total, with 78 final
candidates revised at least once. The final output contains 164
source-attributed proposals. Four module runs retain their discovery output
despite incomplete proposal generation. This configuration
stores one file and line interval per candidate; the general schema
supports multiple locations.

\paragraph{Source selection.}
Appendix~\ref{app:retrieval-data} specifies the source-normalization
procedure and lists the ten pairwise overlaps underlying the mean of 0.1400.

\paragraph{Cost and timing.}
ColPali costs are computed from recorded usage at contracted model rates,
using the token categories specified in Appendix~\ref{app:retrieval-data}. Stage
percentages aggregate the five runs and exclude module-map construction.
Appendix~\ref{app:retrieval-data} gives per-run cost and usage;
Appendix~\ref{app:iocr-details} gives the IOCR timing unit, runtime-ratio
definition and measurement summary.

\clearpage
\section{Candidate-Level Evaluation Data}
\label{app:quality-data}

\subsection{Fabric-X quality scores}
Table~\ref{tab:fabric-scores} gives all 20 scored candidates in evaluation
order. The correctness and severity columns are the two 1--10 ratings used
in Section~\ref{sec:precision}; they are distinct from the score used by the
listwise judge to order candidates. The hit column applies the criterion
in Section~\ref{sec:protocol}. Each row describes the decision or operation examined by a candidate
in the evaluated code snapshot.

\begin{table}[!htbp]
\centering\small
\renewcommand{\arraystretch}{1.18}
\arrayrulecolor{tablerule}
\rowcolors{2}{tablerow}{white}
\begin{tabular}{@{}r p{0.51\linewidth} rrr@{}}
\toprule
\rowcolor{tablehead}
Rank & Decision or operation examined & Correctness & Severity & Hit \\
\midrule
1 & Detecting conflicting signed batches & \cellcolor{scoreteal!30}10 & \cellcolor{scoreteal!30}10 & \textbf{Yes} \\
2 & Comparing batch identities and contents & \cellcolor{scoreteal!30}10 & \cellcolor{scoreteal!15}5 & \textbf{Yes} \\
3 & Verifying participant signatures & \cellcolor{scoreteal!30}10 & \cellcolor{scoreteal!6}2 & No \\
4 & Computing fault-tolerance thresholds & \cellcolor{scoreteal!30}10 & \cellcolor{scoreteal!6}2 & No \\
5 & Validating block-authorization rules & \cellcolor{scoreteal!30}10 & \cellcolor{scoreteal!21}7 & \textbf{Yes} \\
6 & Validating configuration changes & \cellcolor{scoreteal!15}5 & \cellcolor{scoreteal!9}3 & No \\
7 & Parsing complaint messages & \cellcolor{scoreteal!30}10 & \cellcolor{scoreteal!24}8 & \textbf{Yes} \\
8 & Checking certificate validity & \cellcolor{scoreteal!30}10 & \cellcolor{scoreteal!30}10 & \textbf{Yes} \\
9 & Constructing paths from supplied file names & \cellcolor{scoreteal!24}8 & \cellcolor{scoreteal!15}5 & \textbf{Yes} \\
10 & Validating incoming client requests & \cellcolor{scoreteal!24}8 & \cellcolor{scoreteal!15}5 & \textbf{Yes} \\
11 & Checking for previously recorded attestations & \cellcolor{scoreteal!30}10 & \cellcolor{scoreteal!21}7 & \textbf{Yes} \\
12 & Choosing a synchronization target & \cellcolor{scoreteal!3}1 & \cellcolor{scoreteal!3}1 & No \\
13 & Selecting a peer by reported ledger height & \cellcolor{scoreteal!3}1 & \cellcolor{scoreteal!3}1 & No \\
14 & Collecting peer ledger-height responses & \cellcolor{scoreteal!9}3 & \cellcolor{scoreteal!3}1 & No \\
15 & Fetching a batch from multiple peers & \cellcolor{scoreteal!30}10 & \cellcolor{scoreteal!30}10 & \textbf{Yes} \\
16 & Validating consensus control messages & \cellcolor{scoreteal!3}1 & \cellcolor{scoreteal!3}1 & No \\
17 & Rechecking streaming-client authorization & \cellcolor{scoreteal!15}5 & \cellcolor{scoreteal!15}5 & No \\
18 & Selecting an agreed initial ledger block & \cellcolor{scoreteal!3}1 & \cellcolor{scoreteal!3}1 & No \\
19 & Comparing certificate identities & \cellcolor{scoreteal!3}1 & \cellcolor{scoreteal!3}1 & No \\
20 & Configuring server timeouts and limits & \cellcolor{scoreteal!3}1 & \cellcolor{scoreteal!3}1 & No \\
\bottomrule
\end{tabular}
\caption{Complete score data underlying Fabric-X precision.}
\label{tab:fabric-scores}
\end{table}

\subsection{Reference recovery}
Table~\ref{tab:reference-details} expands the aggregate recall results into
individual targets. A reference is counted once even when several output
candidates overlap it; its best rank is the lowest rank among those matches.
We use the matching rule in Section~\ref{sec:protocol}. In llm-d, one missed reference has no candidate in its file; the
other has a candidate in the file but outside the reference's line spans.

\begin{table}[!htbp]
\centering\small
\renewcommand{\arraystretch}{1.18}
\arrayrulecolor{tablerule}
\rowcolors{2}{tablerow}{white}
\begin{tabular}{@{}l p{0.56\linewidth} r@{}}
\toprule
\rowcolor{tablehead}
Subject & Reference target & Best rank \\
\midrule
llm-d & Threshold for separating prefill and decode & 4 \\
llm-d & Weights combining request-routing server scores & 9 \\
llm-d & Timing and ordering of queued request dispatch & 11 \\
llm-d & Block size for indexing reusable prompt computations & 54 \\
llm-d & Selection of a routing strategy for each request & 66 \\
llm-d & Expiry of predicted cache-availability records & No match \\
llm-d & Request assignment across distributed serving nodes & No match \\
vLLM & CPU cache-manager dispatch by eviction policy & 57 \\
WVA & Saturation-based capacity and scaling decisions & 4 \\
\bottomrule
\end{tabular}
\caption{Per-reference recovery. llm-d's request-dispatch target also matches
candidates at ranks 21 and 57; WVA's target also matches ranks 6, 10 and 18.}
\label{tab:reference-details}
\end{table}

The llm-d cache targets concern reusing computations for shared prompt
prefixes. One controls the block size used to index these computations;
the other controls how long the router retains predictions of cache
availability before confirmation. The routing-strategy target selects
which scheduling policy handles a request. The distributed-node target
assigns a request to a serving node using its identifier; an alternative
could account for node load or network conditions.

\section{Repeated Retrieval Runs}
\label{app:retrieval-data}

The five runs use the same versions of ColPali and Spotlights and analyze
the retrieval-engine package. Labels 18a--18e distinguish the repeated
runs in the tables. The supplied objective is to find regions that
most limit retrieval latency and throughput and propose concrete
optimizations. The prompt configurations are identical across these runs.

\subsection{Candidate recurrence}
All 284 candidates have a single file and line interval. For intervals
$[a_1,a_2]$ and $[b_1,b_2]$ in the same file, overlap holds when
$a_1\leq b_2$ and $b_1\leq a_2$. For each candidate occurrence, we count
the runs containing at least one directly overlapping candidate, including
its own run. Table~\ref{tab:recurrence-detail} gives the resulting counts.
This direct-match calculation does not infer a match through intermediate
candidates. The impact-label comparison instead uses connected groups,
as defined in Section~\ref{sec:protocol}.

\begin{table}[!htbp]
\centering\small
\renewcommand{\arraystretch}{1.18}
\arrayrulecolor{tablerule}
\rowcolors{2}{tablerow}{white}
\begin{tabular}{@{}lrrrrrr@{}}
\toprule
\rowcolor{tablehead}
Run & Candidates & In 5 runs & In 4 & In 3 & In 2 & In 1 \\
\midrule
18a & 56 & 42 & 8 & 3 & 2 & 1 \\
18b & 55 & 42 & 6 & 3 & 1 & 3 \\
18c & 61 & 42 & 9 & 6 & 2 & 2 \\
18d & 53 & 41 & 5 & 3 & 2 & 2 \\
18e & 59 & 42 & 9 & 5 & 1 & 2 \\
\midrule
Total & 284 & 209 & 37 & 20 & 8 & 10 \\
\bottomrule
\end{tabular}
\caption{Direct recurrence of candidate occurrences across retrieval runs.}
\label{tab:recurrence-detail}
\end{table}

\subsection{Source overlap}
For each pair of runs, source overlap is the Jaccard ratio of normalized
source sets: intersection size divided by union size. Normalization maps
arXiv page variants and recognized mirrors to a paper identifier, preserves
OpenReview paper identifiers, canonicalizes DOI links, and collapses GitHub
file links across line anchors and revisions to a repository/path key.
Documentation links are normalized at page level. Consequently the measure
concerns source identities, not whether different sources describe the
same technique. Table~\ref{tab:source-pairs} lists all ten pairwise values;
their mean rounds to 0.1400.

\begin{table}[!htbp]
\centering\small
\renewcommand{\arraystretch}{1.18}
\arrayrulecolor{tablerule}
\rowcolors{2}{tablerow}{white}
\begin{tabular}{@{}lr@{}}
\toprule
\rowcolor{tablehead}
Run pair & Source Jaccard \\
\midrule
18a--18b & 0.1387 \\
18a--18c & 0.2066 \\
18a--18d & 0.0915 \\
18a--18e & 0.1333 \\
18b--18c & 0.1476 \\
18b--18d & 0.1394 \\
18b--18e & 0.1620 \\
18c--18d & 0.1262 \\
18c--18e & 0.1395 \\
18d--18e & 0.1151 \\
\bottomrule
\end{tabular}
\caption{Pairwise normalized source overlap.}
\label{tab:source-pairs}
\end{table}

\subsection{Execution measurements}
Table~\ref{tab:run-cost-details} supplies the per-run values underlying the
cost ranges. Tokens include input, output, cache-read and cache-creation
usage; elapsed time is wall-clock time rather than summed agent time.
Costs use contracted model rates. For usage records with a context-specific
Claude model label, the calculation uses the corresponding base-model rate.

\begin{table}[!htbp]
\centering\small
\renewcommand{\arraystretch}{1.18}
\arrayrulecolor{tablerule}
\rowcolors{2}{tablerow}{white}
\begin{tabular}{@{}lrrrr@{}}
\toprule
\rowcolor{tablehead}
Run & Candidates & Tokens (millions) & Hours & Cost (USD) \\
\midrule
18a & 56 & 92.745 & 4.96 & 157.21 \\
18b & 55 & 91.376 & 4.95 & 142.50 \\
18c & 61 & 99.661 & 5.09 & 148.05 \\
18d & 53 & 83.843 & 5.11 & 131.28 \\
18e & 59 & 97.788 & 5.30 & 164.66 \\
\bottomrule
\end{tabular}
\caption{Per-run discovery usage, elapsed time and cost.}
\label{tab:run-cost-details}
\end{table}

\section{IOCR Implementation and Measurement Details}
\label{app:iocr-details}

\subsection{Implementation task}
The task directs program optimization to the word detector's model-loading
and execution configuration. It retains the CPU-only pipeline and requires
character accuracy at least as high as the baseline. The measured region
is steady-state page processing: model construction occurs before the
warm-up and is excluded from timed repetitions. Each repetition sums
page-processing time over the same image set, including detection,
recognition and surrounding pipeline stages.

The baseline passes its detector configuration directly to the runtime
wrapper. It disables the CPU memory arena and memory-pattern planning,
uses sequential graph execution, and sets inter-operation threads from
the available core count. The changed detector derives an effective
configuration before creating its runtime session:

\begin{itemize}
\item Enable the CPU memory arena and extend it by the requested allocation
      size, allowing reuse while controlling memory growth.
\item Keep memory-pattern planning disabled.
\item Set inter-operation threads to one and intra-operation threads to
      the allocated core count (four in this measurement).
\item Disable intra-operation thread spinning.
\end{itemize}

The detector passes these settings to the code that creates its runtime
session. That code is extended to accept per-model overrides, so the
changes apply to word detection while other models retain their defaults.

\subsection{Timing and quality comparison}
Baseline and changed versions execute back-to-back on one node using four
allocated CPU cores, a 32 GB memory allocation, one warm-up and five timed
repetitions each. Let $t_{b,i}$ and $t_{m,i}$ denote full-pass runtimes for
the baseline and modified program. The reported ratio is
\[
 q=\frac{\operatorname{median}_{i=1}^{5}t_{m,i}}
          {\operatorname{median}_{i=1}^{5}t_{b,i}}.
\]
Output quality compares recognized text with reference text, averaging
character accuracy over images. The comparison uses five timed observations per version. Table~\ref{tab:iocr-measurement} gives its
summary statistics. Coefficient of variation (CV) is the standard deviation
divided by the mean, expressed as a percentage.

\begin{table}[!htbp]
\centering\small
\renewcommand{\arraystretch}{1.18}
\arrayrulecolor{tablerule}
\rowcolors{2}{tablerow}{white}
\begin{tabular}{@{}lrr@{}}
\toprule
\rowcolor{tablehead}
Measure & Baseline & Modified \\
\midrule
Timed repetitions & 5 & 5 \\
Median full-pass runtime (seconds) & 150.970 & 135.026 \\
Runtime CV (percent) & 0.66 & 0.91 \\
Mean character accuracy (percent) & 95.53 & 95.53 \\
\bottomrule
\end{tabular}
\caption{IOCR benchmark summary. The runtime ratio is 0.8944, a 10.6\%
reduction, with reported two-standard-error band [0.8818, 0.9070].}
\label{tab:iocr-measurement}
\end{table}

\end{document}